\documentstyle[11pt,dunk2001_asp,twoside,psfig]{article}
\def\edcomment#1{\iffalse\marginpar{\raggedright\sl#1\/}\else\relax\fi}
\reversemarginpar

\def\spose#1{\hbox to 0pt{#1\hss}}
\def\lta{\mathrel{\spose{\lower 3pt\hbox{$\mathchar"218$}}
     \raise 2.0pt\hbox{$\mathchar"13C$}}}
\def\gta{\mathrel{\spose{\lower 3pt\hbox{$\mathchar"218$}}
     \raise 2.0pt\hbox{$\mathchar"13E$}}}

\def\=#1{\overline{#1}}

\def\lvplot{($l,v$) diagram}
\def\lvplots{\lvplot s}

\def\rcr{R_{\rm CR}}

\def\phibar{\varphi_{\rm bar}}

\def\RSUN{R_0}

\def\deg{^\circ}             

\def\kms{{\rm\,km\,s^{-1}}}
\def\kmskpc{{\rm\,km\,s^{-1}\,kpc^{-1}}}

\def\pc{{\rm\,pc}}
\def\kpc{{\rm\,kpc}}

\def\msun{{\rm\,M_\odot}}

\def\etal{{et~al.\ }}
\def\aa #1 #2 {A\&A, #1, #2}
\def\aas #1 #2 {A\&AS, #1, #2}
\def\acm #1 #2 {ACM-Trans Math Software, #1, #2}
\def\ada #1 #2 {Ann Astrophys, #1, #2}
\def\agabstr #1 #2 {Astr Ges Abstr Ser, #1, #2}
\def\aj #1 #2 {AJ, #1, #2}
\def\anach #1 #2 {Astr Nachr, #1, #2}
\def\apj #1 #2 {ApJ, #1, #2}
\def\apjl #1 #2 {ApJL, #1, #2}
\def\apjs #1 #2 {ApJS, #1, #2}
\def\araa #1 #2 {ARAA, #1, #2}
\def\apss #1 #2 {ApSpaceS, #1, #2}
\def\celmech #1 #2 {Cel Mech, #1, #2}
\def\esom #1 #2 {ESO Messenger, #1, #2}
\def\fundcp #1 #2 {FunCosP, #1, #2}
\def\jcp #1 #2 {J Comp Phys, #1, #2}
\def\jfm #1 #2 {J Fluid Mech, #1, #2}
\def\jmp #1 #2 {J Math Phys, #1, #2}
\def\ma #1 #2 {Mitt Astr Ges, #1, #2}
\def\mn #1 #2 {MNRAS, #1, #2}
\def\nat #1 #2 {Nat, #1, #2}
\def\obs #1 #2 {Observatory, #1, #2}
\def\pasj #1 #2 {PASJ, #1, #2}
\def\pasp #1 #2 {PASP, #1, #2}
\def\phyr #1 #2 {PhysRep, #1, #2}
\def\physd #1 #2 {Physica D, #1, #2}
\def\rpp #1 #2 {RepProgPhys, #1, #2}
\def\ssr #1 #2 {Sp Sci Rev, #1, #2}

\begin{document}
\title{The Galactic Bar}
 \author{Ortwin Gerhard}
\affil{Astronomisches Institut, Universit\"at Basel\\ Venusstrasse 7,
CH-4102 Binningen, Switzerland, \\ email: Ortwin.Gerhard@unibas.ch}

\begin{abstract}
  The Milky Way has a barred bulge. This article summarizes the
  current understanding of the main structural parameters and pattern
  speed of the bar, and compares predicted values for the microlensing
  optical depth with the bulge microlensing observations.
\end{abstract}

\section{Introduction}

In the last ten years it has been established that the Milky Way is a
barred Galaxy. More precisely, its central bulge is a rapidly rotating
bar. While there had been a number of papers since the 1970's arguing
that the large non-circular motions seen in HI and CO observations of
the inner Galaxy were best interpreted in terms of a barred potential
(e.g, Peters 1975, Cohen \& Few 1976, Liszt \& Burton 1980, Gerhard \&
Vietri 1986, Mulder \& Liem 1986), it was only in the 1990's that the
combined evidence from the NIR light distribution (Blitz \& Spergel
1991, Weiland \etal 1994, Binney, Gerhard \& Spergel 1997), source
count asymmetries (Nakada \etal 1991, Stanek \etal 1997, Nikolaev \&
Weinberg 1997), gas kinematics (Binney \etal 1991, Englmaier \&
Gerhard 1999, Fux 1999, Weiner \& Sellwood 1999), and large
microlensing optical depth (Udalski \etal 1994, Zhao, Spergel \& Rich
1995, Han \& Gould 1995, Alcock \etal 1997, Stanek \etal 1997) has
convinced most workers in the field.

The emphasis of recent work has shifted towards determining parameters
like the orientation, size, and pattern speed of the bar, and towards
constructing a unifying quantitative model, within which the various
observational results can be coherently explained. Furthermore, from
such a model we should be able to predict the dynamical state of the
Galactic bulge and inner disk, and to understand effects of the bar on
Galactic evolution, such as how it facilitates mass inflow, star
formation, and chemical homogenization of the disk. This work is still
very much in progress, with the answers to several important questions
not yet clear. For example, do we need to make a distinction between
the barred bulge and a bar in the disk, between components of
different age and metallicity?  To what extent do the various tracers
(NIR light, clump giants, NIR star counts and IRAS sources) represent
the mass distribution? Can we quantitatively explain the kinematics of
the Galactic centre gas and the $3 \kpc$ arm, predict the locations of
the Galactic spiral arms, and decide whether they are seen in the old
stars as well as the gas?  How does the large microlensing optical
depth fit in?  Can we correctly predict with dynamical models the
observed stellar kinematics in the bulge and the measured microlensing
duration distribution?

In this article, I would like to report on some steps towards such a
unifying model, concentrating on the photometric structure (\S2) and
pattern speed of the bar (\S3), and the predicted bulge microlensing
(\S4).  For a more extensive review than given here see Gerhard
(1999). All length scales given below are for a Sun-center distance
$\RSUN=8\kpc$.

\section{Photometric structure}

The most detailed models for the distribution of old stars in the
inner Galaxy are currently based on the COBE/DIRBE NIR data.  These
data have complete sky coverage, and broad-band emission maps are
available over a large wavelength range from the NIR to the FIR, but
they have relatively low spatial resolution, they must still be
`cleaned' for residual dust absorption, and they contain no distance
information, so deprojection is not straightforward.  The cleaned data
show that the bulge is brighter and more extended in latitude $b$ at
positive longitudes $l$ than at corresponding $-l$, except for a
region close to the center where the effect is reversed (Weiland \etal
1994, Bissantz \etal 1997). The asymmetry is strongest around
$|l|\simeq10\deg$.  These signatures are as expected for a barred
bulge with its long axis in the first quadrant (Blitz \& Spergel
1991), and contain information about the bar's shape and orientation.
The region of reversed asymmetry at small $|l|$ argues for a bar
rather than a lopsided light distribution; see also Sevenster (1999).

The first models fitted to these data were parametric models assuming
specific functional forms for the luminosity distribution of the
barred bulge and disk and excluding low-latitude regions from the fit
(Dwek \etal 1995, Freudenreich 1998).  Non-parametric models were
constructed by Binney, Gerhard \& Spergel (1997), using a Lucy
algorithm based on the assumption of strict triaxial symmetry, and by
Bissantz \& Gerhard (2002), using a penalized maximum likelihood
algorithm. Both models are derived from the data as cleaned by
Spergel, Malhotra \& Blitz (1997), who used a fully three-dimensional
model of the dust distribution based on the FIR emission data.  The
recent model by Bissantz \& Gerhard includes a spiral arm model as a
prior, maximizing simultaneously eightfold symmetry and smoothness of
the luminosity distribution through corresponding penalty terms in the
likelihood function.  Figure 1 shows two sections through the model.
In this model the bulge-to-disk ratio in NIR luminosity is about
$20\%$, similar to the value given by Kent, Dame \& Fazio (1991).
Other bar and disk properties from the COBE models are summarized
below.  Physical models for the COBE bar can be found for a range of
bar orientation angles, $15\deg \lta \phibar \lta 35\deg$, where
$\phibar$ measures the angle in the Galactic plane between the bar's
major axis at $l>0$ and the Sun-center line.  $\phibar$ must therefore
be determined from other data; see also Zhao (2000).

\begin{figure}[t]
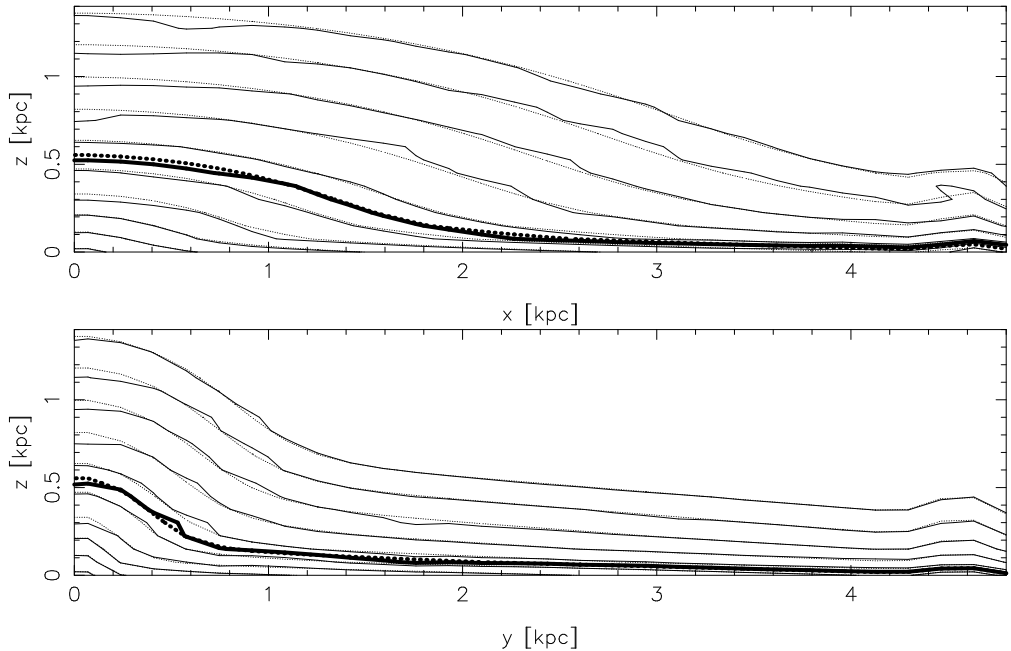

\centerline{\vbox{
\psfig{figure=xz.eps,width=\hsize}
\psfig{figure=yz.eps,width=\hsize}}}
\caption{Top: Section through the luminosity model
  obtained from the COBE L-band data for $\phibar=20\deg$, containing
  the bar's long and short axes.  Bottom: Same, containing the
  intermediate and short axes. After Bissantz \& Gerhard (2002).}
\end{figure}

\begin{figure}[t]
\centerline{\vbox{
\psfig{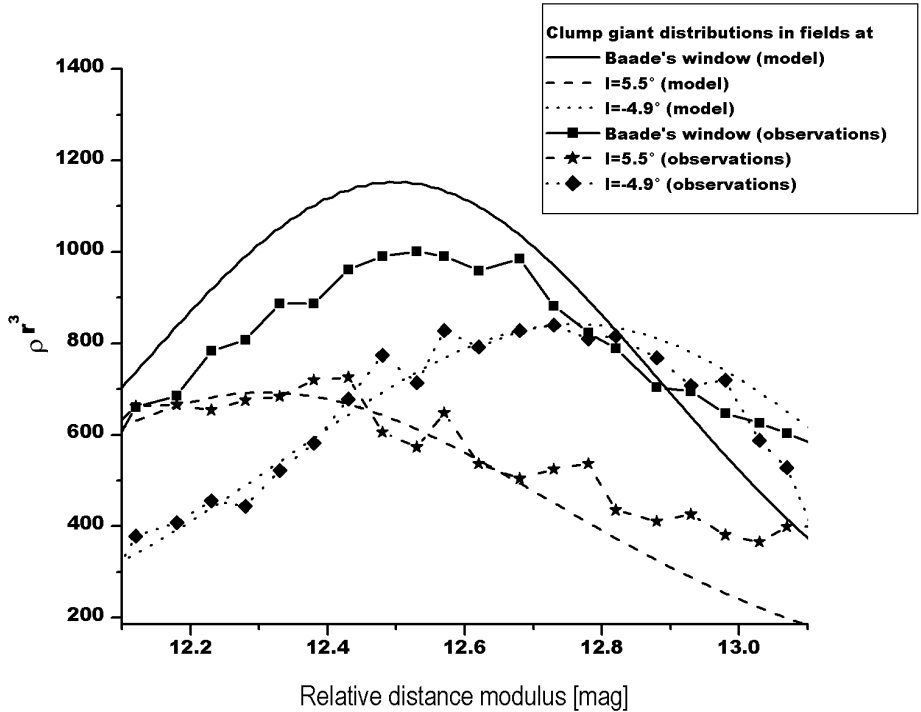}}}
\caption{Apparent magnitude distributions of clump giant stars in
  three fields observed by Stanek \etal (1997). Curves show the
  predictions of the model of Fig.~1 after normalisation and
  convolution with $0.3^m$ intrinsic luminosity spread. From Bissantz
  \& Gerhard (2002).}
\end{figure}

The bar is also seen in starcount observations in inner Galaxy fields.
Stanek \etal (1997) analyzed reddening-corrected apparent magnitude
distributions of clump giant stars in 12 OGLE fields. The small
intrinsic luminosity spread ($\sim$ 0.2-0.3 mag) makes these stars
good distance indicators.  The peak of the distribution is brighter at
$l>0$ where the line-of-sight passes through the near side of the bar.
These fields cover only a small fraction of the sky, but fitting a
parametric model constrains the bar orientation angle as well as the
axis ratios and density profile. Nikolaev \& Weinberg (1997)
reanalyzed the IRAS variable population in a similar spirit; here the
distance information comes from the known range of AGB star
luminosities.  NIR starcounts have also shown longitudinal asymmetries
due to the bar (Unavane \& Gilmore 1998). L\'opez-Corredoira \etal
(1997, 2000) and Hammersley \etal (1999) have modelled the Two Micron
Survey Starcounts (mostly bright K and M giants) in several strips
across the bulge. Structural information on the bulge and disk can be
derived from these data together with a model for the bright-star
luminosity function.  Ongoing work on deeper surveys (ISOGAL, DENIS,
2MASS) will provide important new information on the old stellar
population in the inner Galaxy; first results are given by van Loon
(2001) and Alard (2001), who has interpreted residuals in the 2MASS
star counts from the main bar in terms of a secondary inner bar.

An important goal is to construct a model that simultaneously
reproduces both the starcount and integrated light data. The Bissantz
\& Gerhard (2002) models represent a first step in this direction,
having been constrained a posteriori by the clump giant apparent
magnitude distributions from Stanek \etal (1997).  As Figure 2 shows,
their favoured model matches well the asymmetry of the peak positions
in the fields near $l=\pm 5\deg$ with respect to Baade's window, and
also the relative peak amplitudes within $\sim10\%$, showing that the
line-of-sight distribution of luminosity in the bulge part is
approximately correct.

Modelling the HI and CO \lvplot s provides information on the
gravitational potential of the bar and disk. Several recent gas flow
models (see \S3) have produced \lvplot s with which many features seen
in the observed \lvplot s may be qualitatively understood, such as the
3 kpc arm, the non-circular velocities around the end of the bar, the
cusped-orbit shock transition and inner $x_2$-disk, the molecular ring
and the spiral arm tangent locations, although no model as yet
provides a satisfactory account quantitative account of the entire
observed \lvplots.

The following subsections contain my best summary of the main
bar and disk parameters from this and other work.

{\noindent\bf Bar orientation} From the integrated light alone,
physically reasonable models can be found for $15\deg \lta \phibar
\lta 35\deg$. Starcount models give values between
$\phibar=12\pm6\deg$ (L\'opez-Corredoira \etal 2000) and $20-30\deg$
(Nikoalev \& Weinberg 1997, Stanek \etal 1997).  The models of
Bissantz \& Gerhard (2002) for the DIRBE L-band data, when
additionally constrained by the clump giant apparent magnitude
distributions of Stanek \etal (1997), give an optimal
$\phibar=20-25\deg$, but $\phibar\simeq15-30\deg$ is within the
uncertainties.  The gas-dynamical models and the orbit analysis of
Binney \etal (1991) are also compatible with $15\deg \lta \phibar \lta
35\deg$, depending on whether the emphasis is on the peak in the
terminal velocity curve, the arm morphology, or the magnitude of the
non-circular motions near the 3 kpc arm. Finally, microlensing
observations favour $\phibar\sim15\deg$ (Zhao \& Mao 1996).  Thus a
good working value is $\phibar=20\deg$. Not consistent with this
appear to be the bar model of Hammersley \etal (2000), which is based
on the identification of a region of strong star formation at
$l=27\deg$ with the nearer end of the bar, and the star count results
reported in van Loon (2001), which place the near end of a 1.4 kpc
size bar at negative longitudes.

{\noindent\bf Bar length:} Models based on the DIRBE NIR maps find the
end of the bar around $R_{GC}=3.1-3.5\kpc$, when $\phibar\simeq
20\deg$ (Freudenreich 1998, Binney, Gerhard \& Spergel 1997, Bissantz
\& Gerhard 2002). This is consistent with with the OH/IR stars
(Sevenster 1999), IRAS variables (Nikolaev \& Weinberg 1997), and the
range of $\rcr$ given below, for a fast bar, while other starcount
models use exponential or Gaussian density distributions with shorter
scale-lengths.

{\noindent\bf Bar axis ratios:} The parametric DIRBE models give axial
ratios of about 10:3-4:3. This is in agreement with the new
non-parametric model of Bissantz \& Gerhard (2002), whereas Binney,
Gerhard \& Spergel (1997) had found 10:6:4 without taking into account
the foreground spiral arms. The starcount models give 10:4:3 (Stanek
\etal 1997) and 10:5.4:3.3 (L\'opez-Corredoira \etal 2000).  Thus
there is good overall agreement at around 10:4:3.

{\noindent\bf Disk scale-length:} In the integrated NIR the radial
exponential disk scale $R_D$ is significantly shorter than in the
optical; numerical values are around $2.5\kpc$ (Freudenreich 1998,
Binney, Gerhard \& Spergel 1997) or somewhat shorter ($2.1\kpc$,
Bissantz \& Gerhard 2002).  Hammersley \etal (1999) report
satisfactory agreement of their NIR counts with a model with
$R_d=3.5\kpc$, but have not tested other values of $R_D$.
L\'opez-Corredoira \etal (2000) find that $R_d=3.0\kpc$ is too short
to describe the NIR TMSS counts well. Earlier starcount models (Robin
\etal 1992, Ortiz \& L\'epine 1993) favour a short disk scale,
$R_d=2.5\kpc$. It is not clear what causes the differences between the
various starcount models, and between the TMSS starcounts and the
integrated NIR light.  There may be some interplay between the disk
scale-length, the bulge profile, and the spiral arm luminosity
distribution. To clarify this will need further work with spatially 
complete data.

\section{Pattern speed}

The pattern speed $\Omega_p$, or equivalently, the corotation radius
$\rcr$, is the most important dynamical parameter of the bar, because
it determines the orbits of stars in the disk. Bar pattern speeds can
be parametrized by the ratio $R=D_L/a_b$, where $D_L$ is the
Lagrangian radius at which the gravitational and centrifugal forces
cancel out, approximately equal to the axisymmetric corotation radius,
and $a_b$ the bar's semi-major axis length. The Milky Way bar is a
fast bar ($R\gta 1$), as are the bars in external galaxies for which
pattern speeds have been determined (see Debattista, Gerhard \&
Sevenster 2002 for a summary). Three methods have been used to
estimate the pattern speed of the Galactic bar:

{\noindent \bf Hydrodynamical simulations:} The gas-dynamical
simulations of Englmaier \& Gerhard (1999) and Fux (1999) agree in
their interpretation of the 3 kpc arm as one of the lateral arms close
to the bar, placing it inside corotation.  Sevenster (1999) argues
that the 3 kpc arm is part of an inner ring, which would also place it
slightly inside the corotation radius $\rcr$.  The main Galactic
spiral arms outside $\rcr$, on the other hand, imply an upper limit
for $\rcr$, but this is more model-dependent. These gas-dynamical
models thus give a range of $3\kpc \lta \rcr \lta 4.5\kpc$ (see
Gerhard 1999 where also a corresponding resonance diagram is shown).
Weiner \& Sellwood (1999) have concentrated in their models to
reproduce the extreme (non-circular) velocity contour in the HI data
of Liszt \& Burton (1980), using gravitational potentials that include
a Ferrers bar.  They favour a somewhat lower value for $\Omega_p\simeq
42\kmskpc$, corresponding to $\rcr\simeq 5.0\kpc$, but also need a
significantly larger bar angle $\phibar\simeq 34\deg$ than favoured by
the results reported in \S2. The recent SPH simulations of Bissantz,
Englmaier \& Gerhard (2002) based on the COBE bar potential of
Bissantz \& Gerhard (2002) find $\Omega_p=60\pm5\kmskpc$,
corresponding to $\rcr=3.3\pm0.3\kpc$. The best of these models
include a second, lower pattern speed for the Galactic spiral arms.
These models give a very good representation of the \lvplot\ in the
region outside the bar, but still underpredict the non-circular
velocities in the 3 kpc arm. However, both Bissantz \etal (2002) and
Fux (1999) find that the amplitude of these non-circular velocities is
influenced by additional parameters such as the details and
time-dependence of the gravitational potential in the bar-spiral
transition region, which are almost certainly not reliably modelled in
the simulations.

{\noindent \bf Orbital resonances:} Dehnen (2000) has interpreted
features in the local stellar velocity distribution as due to the
outer Lindblad resonance with the bar, resulting in $\rcr =0.55\pm0.05
R_0 \simeq 4.4\pm0.4\kpc$ for $R_0=8\kpc$, near the upper end of the
range from the gas-dynamical models, and corresponding to
$\Omega_p=51\pm4 \kmskpc$. While the match to the Hipparchos data
appears convincing, this analysis relies on the assumption that the
quadrupole moment of the bar is strong enough near the Sun to dominate
the resonance, as compared to that of the spiral arms.  In the new
COBE models of Bissantz \& Gerhard (2002) the more elongated bar as
compared to previous models does dominate the quadrupole moment near
the Sun, but there still is the caveat that we know little about the
amplitude of the Galactic spiral arms in the old stellar component.

{\noindent \bf Direct method:} Recently, Debattista \etal (2002) have
adapted the Tremaine-Weinberg (1984) method to a sample of tracers in
the Milky Way, and have used it to analyze the sample of OH/IR stars
collected by Sevenster and collaborators (see Sevenster \etal 2001).
For $R_0=8\kpc$ and $V_0=220\kms$ and assuming inward LSR motion of
$u_{\rm LSR}=0$ as suggested by HI absorption observations
(Radhakrishnan \& Sarma 1980), they find $\Omega_p=59\pm5\kmskpc$ for
these sources with a possible additional systematic error of perhaps
$10\kmskpc$. The OH/IR stars which carry the main signal in the
analysis are at low latitude and slightly outside the region of the
bar, according to the models described in \S2. Thus the feature
responsible for the measured pattern speed most likely is the inner
parts of the spiral arms, particularly the Scutum arm tangent. The
high value of $\Omega_p$ suggests that this spiral arm might be
coupled to the bar; perhaps it is even an inner ring rather than a
spiral arm. Such rings are common in barred galaxies (Buta 1995), and
being elongated along the bar, corotate with the bars they contain.
Sevenster \& Kalnajs (2001) have postulated such a ring in the Milky
Way. In either case, the pattern speed of the outer spiral arms could
be lower, as favoured by the most recent hydrodynamic simulations and
the independent work of L\'epine \etal (2001) on Cepheid velocities.

Taking these results together, a good estimate for the corotation
radius is $\rcr=4\pm 0.5\kpc$. Recall from \S2 that the major axis
length of the COBE bar is approximately $3.3\pm0.2\kpc$, so $R\simeq
1.2\pm0.2$.

\section{Microlensing}

Microlensing observations provide important new constraints on the
Galactic mass distribution.  Several hundred microlensing events have
been observed towards the Galactic bulge by the OGLE and MACHO
collaborations. These observations give information about the
integrated mass density towards the survey fields as well as about the
lens mass distribution. The most robust observable is the total
optical depth averaged over the observed fields, $\tau$.  Early
measurements gave surprisingly high values $\tau_{-6}\simeq 2-4$
(Udalski \etal 1994, Alcock \etal 1997), where
$\tau_{-6}\equiv\tau/10^{-6}$.  Alcock et al.\ (2000a) determined
$\tau_{-6}=2.43^{+0.39}_{-0.38}$ for all source types from 99 events
centered on $(l,b) = (2.68\deg, -3.35\deg)$, using a difference image
analysis (DIA) technique.  From this measurement they deduced an
optical depth $\tau=(3.23\pm0.5)\times10^{-6}$ for bulge sources only
in the same field.  Popowski \etal (2000) published a preliminary
analysis of 52 clump giant sources in 77 Macho fields, resulting in a
lower $\tau_{-6}=2.0\pm0.4$ centered on $(l,b)=(3.9\deg,-3.8\deg)$.
The important advantage of using clump giant sources is that they do
not suffer from blending problems.

The measured optical depth values in these bulge fields are
unexpectedly high. Axisymmetric Galactic models predict
$\tau_{-6}\simeq 1-1.2$, insufficient to explain the quoted values
(Kiraga \& Paczynski 1994, Evans 1994).  Models with a nearly end--on
bar enhance $\tau$ because of the longer average line-of-sight from
lens to source.  The maximum effect occurs for $\phi\simeq
\arctan(b/a)$ when $\tau_{\rm bar}/\tau_{\rm axi}\simeq
(\sin2\phi)^{-1}\simeq 2$ for $\phi=15\deg$ (Zhao \& Mao 1996).
$\tau$ is also proportional to the mass of the bar/bulge and increases
with the bar length.

Nonetheless, models based on barred mass distributions derived from Milky
Way observations typically give only $\tau_{-6}\simeq 1-2$ (e.g., Zhao,
Spergel \& Rich 1995, Stanek \etal 1997, Bissantz \etal 1997),
significantly less than most of the measured optical depths.  Figure 3
shows predicted optical depths for the new COBE bar model of
Bissantz \& Gerhard (2002) as a function of latitude, at the central
longitude positions of the newest microlensing measurements.  The mass
normalization of the disk and bulge in this model is calibrated by
assuming constant L-band mass-to-light ratio and by matching the
predicted gas flow velocities in a hydrodynamic simulation to the
Galactic terminal velocity curve; see above. The numerical values
are $\tau_{-6}=1.1$ for all sources at the position of the DIA
measurement and $\tau_{-6}=1.27$ for clump giant sources at the
centroid position given by Popowski \etal (2000).  

The model prediction for clump giant sources is within $1.8\sigma$ of
the Popowski \etal MACHO value. On the other hand, the recent DIA
value is still some $3.2\sigma$ away from the model prediction.
Because the apparent magnitude distributions for clump giant stars
predicted by this model agree closely with those measured by Stanek
\etal (1997) -- see Fig.~2 -- this model gives a good approximation to
the distribution of microlensing {\sl sources}.  Changing the quoted
optical depths substantially is therefore hard unless the mass
distribution of the {\sl lenses} differs substantially from that of
the sources. While the NIR model prediction could be slightly
increased if the mass-to-light ratio were not spatially constant, this
is only a $\sim 20\%$ effect since limited by the terminal velocity
curve (Bissantz \etal 1997).

\begin{figure}
\centerline{\vbox{
\psfig{figure=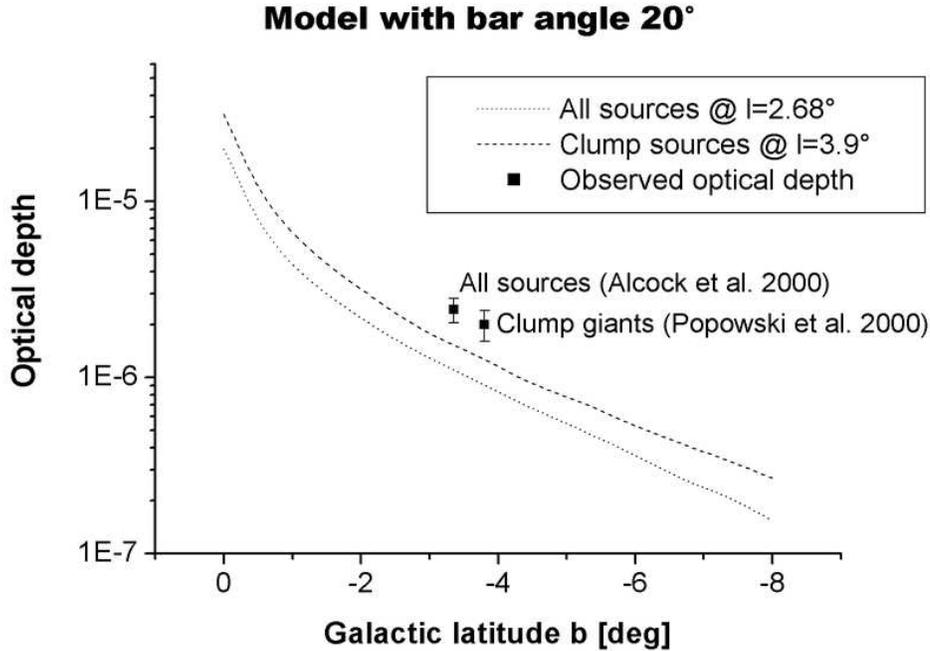,width=\hsize}}}
\caption{Microlensing optical depths for the model of Fig.~1 at the
  longitudes of the newly published MACHO results, plotted as a
  function of galactic latitude. The points with error bars show the
  observed optical depths. The upper curve is for clump giant sources,
  the lower curve for all sources using a simple parametrization of
  the magnitude cut-off. From Bissantz \& Gerhard (2002). }
\end{figure}

General arguments (Binney, Bissantz \& Gerhard 2000) show that an
optical depth for bulge sources as large as that implied by the MACHO
DIA measurement is very difficult to reconcile with the Galactic
rotation curve and local mass density, even for a barred model and
independent of whether mass follows light.  To illustrate this, the
extra optical depth required on top of the Bissantz \& Gerhard (2002)
model prediction would correspond to an additional mass surface
density towards the bulge of some $1500 \msun/\pc^2$ at the optimal
location half-way to the bulge. This is comparable to the luminous
surface mass density in the NIR model (some $3600 \msun/\pc^2$ but not
optimally located.  Because the model predictions are fairly robust,
it is important to check again whether the DIA measurement could still
be significantly affected by blending.

These results have a further important implication.  Because a model
based on the maximal disk assumption and calibrated with the terminal
velocities still underpredicts the observed bulge microlensing optical
depths, the contribution of a non-lensing CDM dark halo to the mass
distribution in the inner Galaxy cannot be large (from the LMC
microlensing experiments of Alcock \etal 2000b at most a fraction of
the dark matter halo can contribute to microlensing.) Thus the bulge
microlensing results argue strongly for a massive disk and low central
density halo (see also Binney \& Evans 2001). This is consistent
with the fact that the NIR disk with constant mass-to-light ratio
fitted to the terminal velocity curve correctly predicts the local
disk surface mass density, leaving little room for extra mass in the
inner Galaxy (Gerhard 1999).

\end{document}